\documentclass[aps,prl,twocolumn,superscriptaddress,amsmath,amssymb,floatfix]{revtex4-1}

\usepackage{times,mathptmx}
\usepackage{graphicx,pstricks,subfigure}
\usepackage{epstopdf}
\usepackage{bm}

\begin{document}

\title{Threshold resonance effects in reactive processes}

\author{I. Simbotin}
\affiliation{Department of Physics, University of Connecticut, 2152
Hillside Rd., Storrs, CT 06269, USA}

\author{S. Ghosal}
\affiliation{Department of Physics, University of Connecticut, 2152
Hillside Rd., Storrs, CT 06269, USA}
\affiliation{Department of Chemistry,
  Birla Institute of Technology and Science Pilani,
  Hyderabad Campus, Jawahar Nagar, Shameerpet (M), Hyderabad, 500078, India}

\author{R. C\^ot\'e}
\affiliation{Department of Physics, University of Connecticut, 2152
Hillside Rd., Storrs, CT 06269, USA}
\affiliation{Institute for Quantum Computing, University of Waterloo, Waterloo, 
Ontario, Canada N2L 3G1}

\date{\today}

\begin{abstract}
  We investigate the effect of near threshold resonances in reactive
  scattering at low energy.  We find a general type of anomalous
  behavior of the cross sections, and illustrate it with a real system
  (H$_2$ + Cl).  For inelastic processes, the anomalous energy
  dependence of the total cross sections is given by
  $\sigma\sim\varepsilon^{-3/2}$.  The standard threshold behavior
  given by Wigner's law ($\sigma\sim\varepsilon^{-1/2}$) is eventually
  recovered at vanishing energies, but its validity is now limited to
  a much narrower range of energies.  The universal anomalous behavior
  leads to reaction rate coefficients behaving as $K\sim 1/T$ at low
  temperatures, instead of the expected constant rate of the Wigner
  regime.  We also provide analytical expressions for $s$-wave cross
  sections, and discuss the implication in ultracold physics and
  chemistry.
\end{abstract}



\maketitle

In recent years, the level of control over the interaction in
ultracold gases, {\it e.g.}, by using magnetically tuned Feshbach
resonances \cite{RMP-FR} or by orienting ultracold molecules
\cite{paper-JILA,Jason-PRL}, has allowed the investigation of various
phenomena in degenerate quantum gases ({\it e.g.}, BEC-BCS cross-over,
solitons, multi-component condensates, etc.)
\cite{RMP-bose,RMP-fermi},
as well as of exotic three-body Efimov states \cite{efimov}.  The
advances in the formation of cold molecules
\cite{carr2009,dulieu2011,mol-papers} are paving the way to the study
\cite{paper-JILA,sawyer2011} and control \cite{quemener2012} of cold
chemical reactions.  In many of these studies, resonances near the
scattering threshold are a key ingredient, as recently explored
experimentally \cite{h2co-prl-2012,shd-prl-2012,narevicius-2012} and
theoretically \cite{Bohn:PRA87:2013} in chemistry of low temperature
systems.

In this Letter, we explore the effect of near threshold resonances
(NTR) in the entrance channel of a reactive scattering system.  In the
absence of NTR, it is well known that scattering cross sections behave
at ultralow energy $\varepsilon$ according to Wigner's threshold law
\cite{wigner}; namely, the elastic cross section tends to a constant,
while the inelastic cross sections behave as $\varepsilon^{-1/2}$,
when $\varepsilon\rightarrow 0$.  However, Wigner cautioned that, when
resonance poles are present near channel thresholds, renewed attention
should be paid to the low energy behavior of cross sections
\cite{wigner,hossein}.  Indeed, resonances in low energy collisions,
originally analyzed by Bethe~\cite{Bethe:PR:1935} for collisions
between neutrons and nuclei, and by Fano in atomic collisions
\cite{fano}, may affect the threshold behavior.

\begin{figure}[b]
\includegraphics[clip,width=1.0\linewidth]{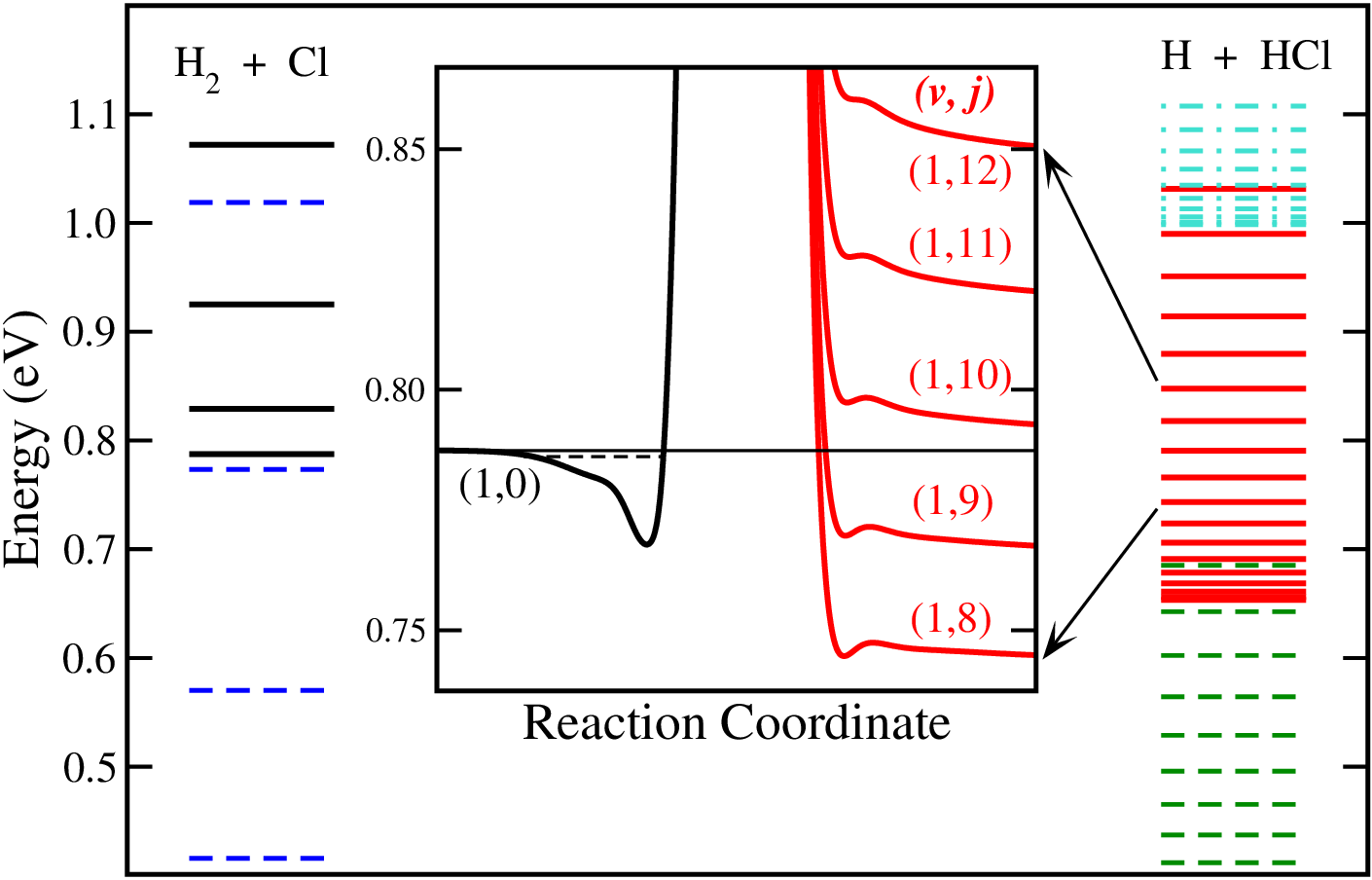}
\caption{Rovibrational energy levels for H$_2$ and HCl.  For H$_2$ we
  show the rotational series for $v=0$ (dashed blue) and $v=1$ (solid black).
  Similarly for HCl, we have $v=0$ (dashed green), $v=1$ (solid red) and $v=2$
  (dot-dashed cyan).  The inset shows the diagonal matrix elements of the
  potential energy surface for the rovibrational channels that are
  near the entrance channel; the dashed line represents a weakly
  quasi-bound level for the van der Waals complex Cl$\cdots$H$_2$ in
  the entrance channel H$_2(1,0)$.  }
\label{fig:h2cl-PES}
\end{figure}

For our study, we selected a barrier-dominated reaction containing a
large number of channels, namely H$_2$ + Cl $\rightarrow$ H + HCl.  In
particular, for H$_2$ in the initial ro-vibrational level
($v$=1,$j$=0), the shallow potential well of the Cl$\cdots$H$_2$
van~der Waals complex supports quasibound states, which can lead to
resonances in the entrance arrangement (see Fig.~\ref{fig:h2cl-PES})
\cite{ClH2:Science:2008}; note that there are no nearby closed
channels that could give Feshbach resonances. This system was recently
investigated at ultralow temperatures by Balakrishnan
\cite{Bala:H2Cl:2012}.  To explore the effect of near threshold
resonances (NTR) on reactive scattering, we vary the mass of H, as was
done in~\cite{Bodo:JPB:2004}, so that the channel thresholds in both
arrangements H$_2$--Cl and H--HCl shift at different rate, an approach
similar to modifying the potential surface itself~\cite{JMH:JCP:2007}.
The scattering cross section from an initial internal state $i$ to a
final state $f$ is given by \cite{alex-bala}
\begin{equation}
   \sigma_{i\rightarrow f}(\varepsilon_i) = \frac{\pi}{k_i^2}
   \sum_{J=0}^{\infty} \left( \frac{2J+1}{2j+1}\right) \sum_\ell
   \sum_{\ell'} \left| \delta_{if} - S_{if}^J\right|^2 \; ,
   \label{eq:full-xsection}
\end{equation}
where $\ell = |J-j|, \dots, J+j$ and $\ell'=|J-j'|,\dots,J+j'$; $\pmb
J = \pmb j + \pmb\ell = \pmb j' + \pmb\ell'$ is the total angular
momentum, with molecular rotational momentum $\pmb j$ and orbital
angular $\pmb \ell$ in the entrance channel $i$, and corresponding
quantum numbers $J$, $j$, and $\ell$ (the primes indicate the exit
channel $f$). Here, $\varepsilon_i = \hbar^2 k_i^2/2\mu$ is the
kinetic energy with respect to the entrance channel threshold, $k_i$
the wave number, and
$\mu^{-1}=m_{\mathrm{H}_2}^{-1}+m_{\mathrm{Cl}}^{-1}$ the reduced mass
in the entrance arrangement.  We are focusing on the effect of
resonances at ultralow temperatures, and thus consider only $s$-wave
scattering with $\ell =0$, which requires $J=j$ and thus
$(2J+1)/(2j+1)=1$. In addition, we limit ourselves to molecules
initially in their rotational ground state with $j=0$, so that $J=0$,
and thus $\ell'=j'$.  This simplifies Eq. (\ref{eq:full-xsection}) to
\begin{equation} \sigma_{i\rightarrow f}(\varepsilon_i) = \frac{\pi}{k_i^2}
    \left| \delta_{if} - S_{if}^{J=0}(k_i)\right|^2 \; . 
    \label{eq:sigma-s-wave}
\end{equation}
We note that with $\ell'=j'$, the centrifugal term in the exit
channels becomes large for high rotational diatomic levels and makes
the van~der Waals potential well disappear (see
Fig.~\ref{fig:h2cl-PES}).  For the sake of clarity, we omit the
subscript $i$ in $\varepsilon_i$ and $k_i$, and $J$ in $S_{if}^{J=0}$,
in the rest of this Letter.

In the zero-energy limit, the elastic and total inelastic cross sections
are described in terms of the complex scattering length \cite{bala-cplett}
$a =\alpha -i\beta$ for a given entrance channel $i$,
\begin{equation}
\begin{array}{lllll}
   \sigma^{\rm in.} (\varepsilon ) & \equiv & \displaystyle \sum_{f\neq i} 
   \sigma_{i\rightarrow f}(\varepsilon)
   & \rightarrow & \displaystyle \frac{4\pi}{k} \beta\;, \\ 
      \sigma^{\rm el.} (\varepsilon ) & \equiv & \sigma_{i\rightarrow i}(\varepsilon) 
   & \rightarrow & 4\pi (\alpha^2 + \beta^2) \;. 
\end{array}
 \label{eq:alpha-beta}
\end{equation}
Although these limits remain valid in the presence of NTR, their applicability
will be limited to a much narrower domain of energies.  For the remainder of
the low energy regime, a new behavior emerges, as we now show.

\begin{figure}[t]
\includegraphics[clip,width=0.9\linewidth]{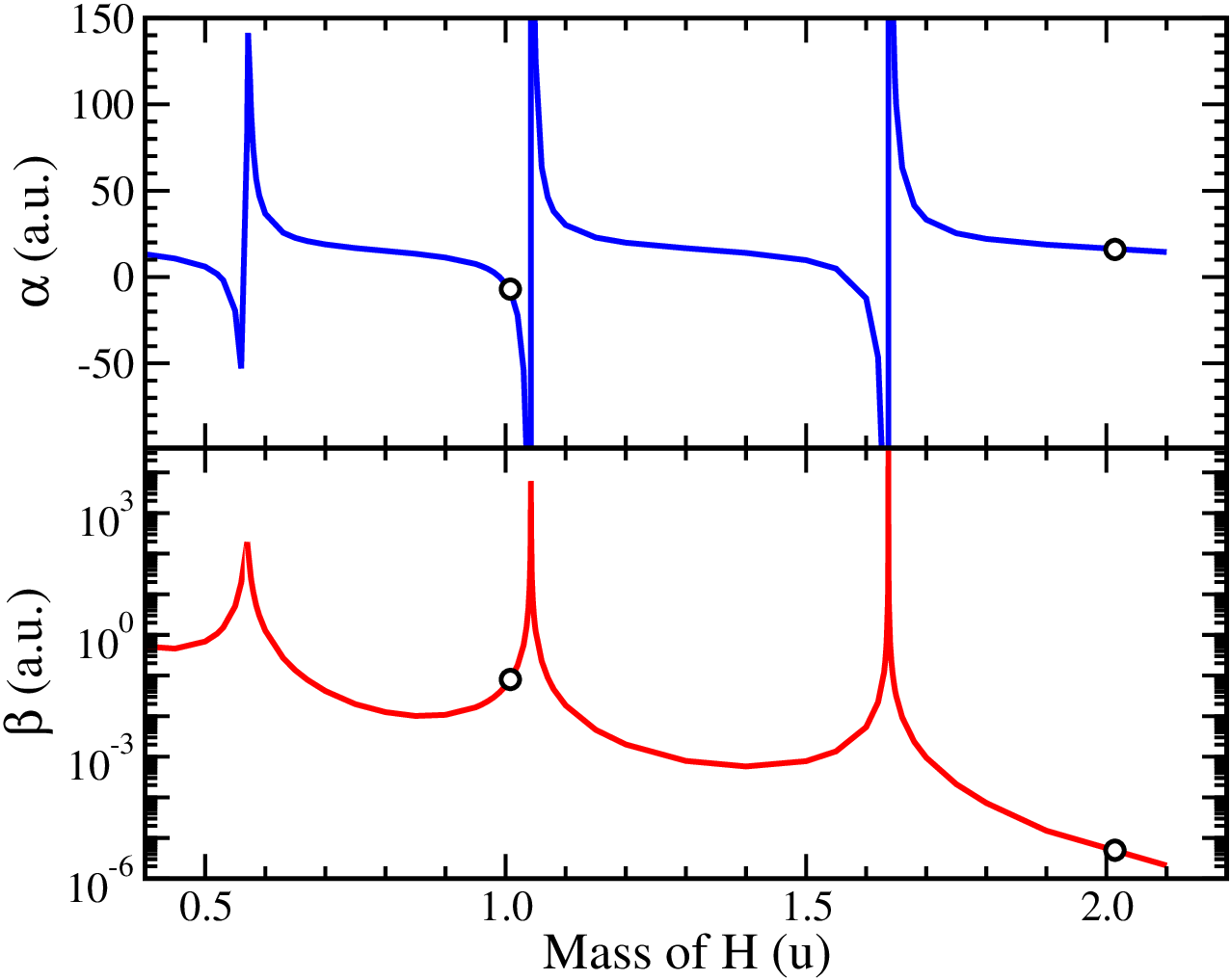}
\caption{Real ($\alpha$) and imaginary ($\beta$) components of the
  scattering length as functions of the mass of the hydrogen atom $m$.
  The true masses of hydrogen and deuterium are indicated by circles.
  We extracted the value of $\alpha$ from the complex phase of the
  diagonal matrix element $S_{ii}$ (see text) and $\beta$ from the
  total inelastic cross section, at the lowest collision energy used
  in our computations, $E/k_B=1$\,nK.}
\label{fig:alpha-beta}
\end{figure}
\begin{figure}[t]
\includegraphics[clip,width=1.0\linewidth]{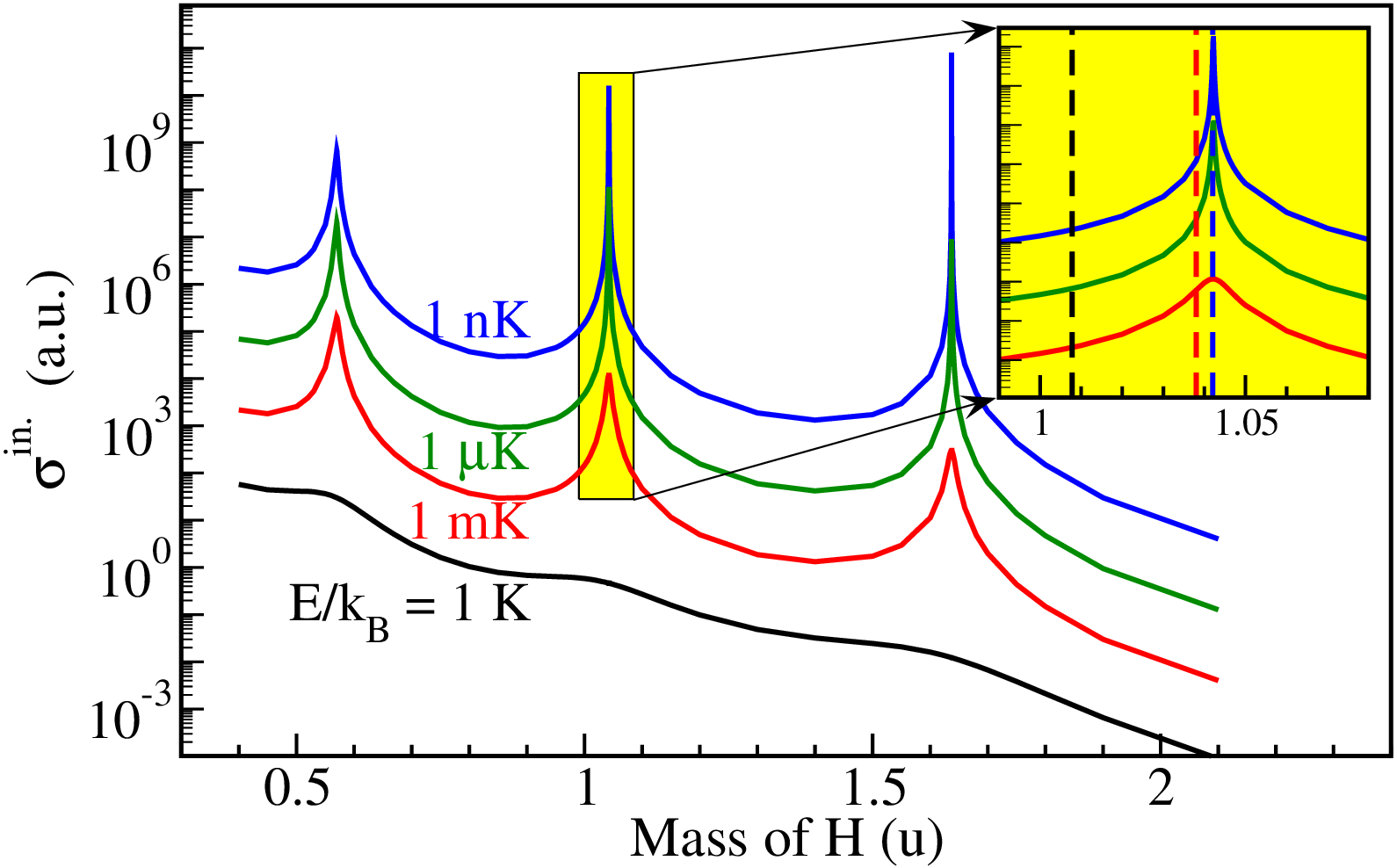}
\caption{Total inelastic cross section as a function of mass $m$ of H
  for different scattering energies in units of Kelvin. The inset
  shows $\sigma^{\rm in.}$ in the vicinity of the resonance with the
  vertical dashed lines corresponding to
  $m=1.0078\,\mathrm{u}=m_{\rm{H}}$ (true mass, in black), 1.038\,u
  (red), and 1.042\,u (blue), from left to right respectively.}
\label{fig:cross-sections}
\end{figure}

The results we present here were obtained using the \textsc{abc}
reactive scattering code of Manolopoulos and
coworkers~\cite{ABC:CPC:2000}, which we have optimized for ultralow
energies.  We used the potential energy surface developed by Bain and
Werner~\cite{BWpes:JCP:2000}.  Fig.~\ref{fig:alpha-beta} shows the
variation of the real and imaginary parts of the scattering length for
the entrance channel H$_2(v=1,j=0)$+Cl, as we scan the mass $m$ of
H. The open circles indicate the true masses of H and D; H lies on the
wing of a resonance in the entrance channel, which gives a sharp
increase for $\beta$. In Fig.~\ref{fig:cross-sections}, we show the
total inelastic cross section (including both quenching and reaction)
as a function of $m$ for a few energies (in Kelvin).  From
Eq.~(\ref{eq:alpha-beta}), we expect $\sigma^{\rm in.}$ to simply
scale as $\varepsilon^{-1/2}$, {\it i.e.}, showing as equidistant
curves for the energies chosen in the logarithmic scale; this is
indeed the case, except in the vicinity of the resonances. In the
inset, we zoom on the resonance, with three masses $m$ of
increasing values being singled out (dashed vertical lines)
corresponding to 1.0078\,u$=m_{\rm H}$ (true mass), 1.038\,u, and
1.042\,u, respectively; the inset indicates that the energy dependence
follows Wigner's threshold law only for $m_{\rm H}$, and departs from
it as $m$ approaches the resonance.  Thus, we focus our attention on
these three masses, and we analyze in detail the energy dependence of
the cross sections.

\begin{figure}[t]
\includegraphics[clip,width=.474\linewidth]{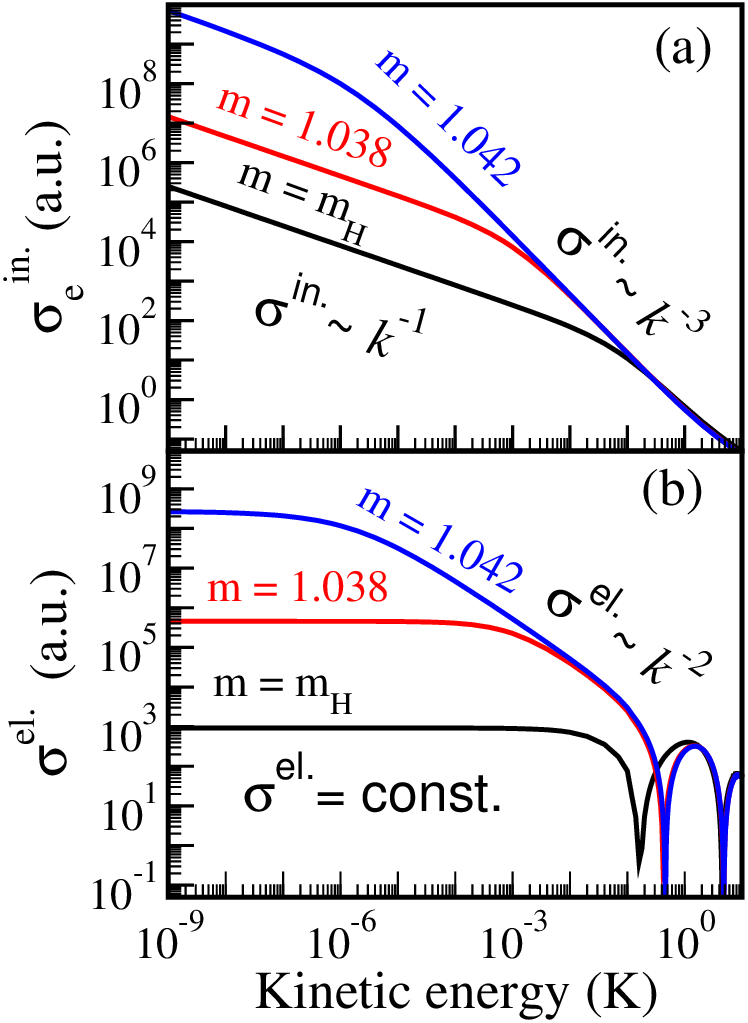}
\includegraphics[clip,width=.5\linewidth]{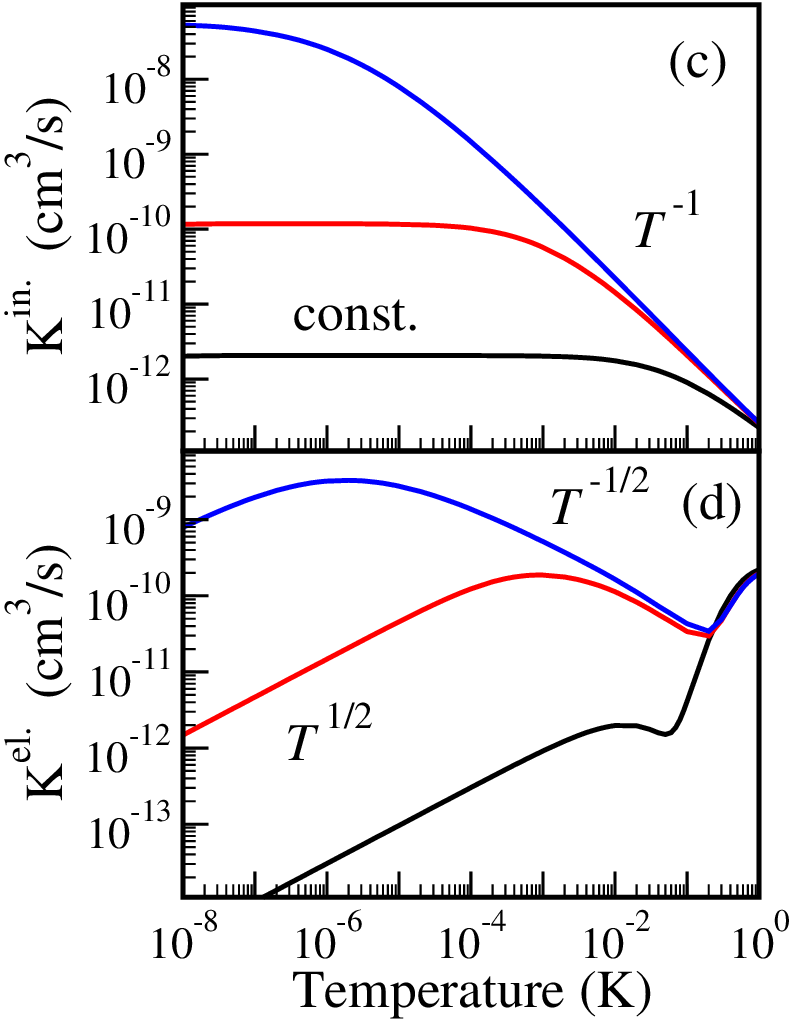}
\caption{\label{fig:rate} Energy dependence of the reaction cross 
  section (a) and elastic cross section (b).  Temperature dependence of 
  rate coefficients for reaction (c) and elastic scattering (d).  The results 
  shown are for the entrance channel  H$_2(v\!\!=\!\!1,\,j\!\!=\!0)$ + Cl
  $\rightarrow$ H + HCl, and they correspond to different masses of
  H, as indicated for each curve.}
\end{figure}

In Fig.~\ref{fig:rate}(a), we show the low energy behavior of 
$\sigma^{\rm in.}$ for the three masses
selected above.  When $\varepsilon\rightarrow 0$, $\sigma^{\rm in.}$
reaches the Wigner's regime, scaling as $\varepsilon^{-1/2}$ for all
three masses. However, for masses closer to the resonance shown in
Fig.~\ref{fig:cross-sections}, the scaling changes to
$\varepsilon^{-3/2}$.  In addition, the behavior appears to be
universal, {\it i.e.}, as $m$ nears the resonance, $\sigma^{\rm in.}$
has the same value until it deviates from the universal NTR
$\varepsilon^{-3/2}$ scaling to join the Wigner $\varepsilon^{-1/2}$
scaling at lower collision energies.  For completeness sake, we also
show the elastic cross sections $\sigma^{\rm el.}$ for the same masses
in Fig.~\ref{fig:rate}(b); the Wigner regime's constant cross section
as $\varepsilon\rightarrow 0$ changes to the expected
$\varepsilon^{-1}$ scaling for $m$ near a resonance.  Finally,
Figs.~\ref{fig:rate}(c) and (d) depict the corresponding rate
coefficients $K^{\rm in.}(T)$ and $K^{\rm el.}(T)$ obtained from
thermal averaging with a Maxwell distribution of relative velocity 
$v_{\rm rel.}$ characterized
by the temperature $T$, i.e., $K(T)=\langle v_{\rm rel.} \sigma \rangle_T$.
Again for $m$ close to the resonance, there is a significant
enhancement of $K^{\rm in.}$ scaling as $T^{-1}$ until $T$ is small
enough that the Wigner regime is reached and $K^{\rm in.}$ becomes
constant. The corresponding scaling for $K^{\rm el.}$ changes from
$T^{-1/2}$ for NTR to $T^{1/2}$ for Wigner's regime. This change in
behavior is due to the fact that, for $m\approx1.042$\,u, the potential
well of the Cl$\cdots$H$_2$ van~der~Waals complex is acquiring a new
$s$-wave bound state; this so-called zero energy bound state is
responsible for the large increase of the probability amplitude at
short range, where the reaction takes place.

This anomalous behavior (namely, an abrupt increase followed by a
gradual transition into the Wigner regime as the energy decreases) is
due to the presence of a resonance pole near the threshold of the
entrance channel. To understand it, we focus our attention on the
S-matrix. Recalling that $\sum_f |S_{if}|^2 = |S_{ii}|^2 + \sum_{f\neq
  i} |S_{if}|^2 =1$, and using Eq.~(\ref{eq:sigma-s-wave}), we rewrite
the cross sections in Eq.~(\ref{eq:alpha-beta}) in terms of the
diagonal element $S_{ii}(k)$;
\begin{eqnarray}
   \sigma^{\rm in.}(\varepsilon ) & = & \frac{\pi}{k^2} \left(1-|S_{ii}(k)|^2\right),
   \label{eq:sigma-in-S-matrix} \\
   \sigma^{\rm el.}(\varepsilon ) & = & \frac{\pi}{k^2} |1-S_{ii}(k)|^2.
   \label{eq:sigma-el-S-matrix}
\end{eqnarray}
The analytical properties of the $S$-matrix for single-channel $s$-wave 
scattering follow from those of the Jost function ${\cal J}(k)$;
\begin{equation}
   S(k) = \frac{{\cal J}(k)}{{\cal J}(-k)} \; .
\end{equation}
Specifically, if $k=z$ is a zero of the Jost function (and of the $S$-matrix),
then $k=-z$ is a pole of the of the $S$-matrix.  Hence, given a resonance
in the entrance channel $i$, we rewrite the diagonal $S$-matrix element
as a product of a resonant part (due to a zero of $S$ located at $k=z$)
and a non-resonant part $\tilde{S}_{ii}(k)$
\begin{equation}
   S_{ii}(k) = \frac{z-k}{z+k}\tilde{S}_{ii}(k) \;.
   \label{eq:S_ee}
\end{equation}
The background contribution $\tilde{S}_{ii}(k)$ is assumed to be a
slowly varying function of $k$, and can be described as
$\tilde{S}_{ii}(k)=e^{2i\tilde\delta_i}$, where the background
phaseshift $\tilde{\delta}_i$ is implicitly for the $\ell=0$ partial
wave.  Here, we are interested in the case when $|z|$ is very small,
such that the energy associated with the pole,
$|E_z|=\hbar^2|z|^2/2\mu$, is within (or near) the ultracold domain.

Writing the background scattering length as
$\tilde{a}_i = \tilde{\alpha}_i-i\tilde{\beta}_i$, and
assuming that $\tilde{\delta}_i \approx -k \tilde{a}_i$ as $k\rightarrow 0$,
we get
\begin{equation}
\label{eq:S-tilde}
   \tilde{S}_{ii}(k) = e^{-2ik\tilde{a}_i} = e^{-2ik\tilde{\alpha}_i}e^{-2k\tilde{\beta}_i}\;,
\end{equation}
so that $|\tilde{S}_{ii}(k)|^2=e^{-4k\tilde{\beta}_i}$. Expressing the
zero in the complex plane as $z\equiv z'-iz''$, then $|S_{ii}(k)|^2 =
e^{-4k\tilde{\beta}_i} |z-k|^2/|z+k|^2$ in
Eq.~(\ref{eq:sigma-in-S-matrix}) leads to
\begin{eqnarray}
   \sigma^{\rm in.} & = & \frac{\pi}{k^2} \frac{e^{-2k\tilde{\beta}}}{|z+k|^2} 
   \left\{ 2 (k^2 +|z|^2) \sinh (2k\tilde{\beta}) \right. \nonumber \\
   & & \makebox[.75in]{ } \left. {}+4 k z' \cosh (2k\tilde{\beta}) \right\} \;,
   \label{eq:sigma_in-full}
\end{eqnarray}
where we omitted the channel subscript.  As $k\rightarrow 0$, we recover the
two different behaviors shown in Fig.~\ref{fig:rate}(a).  Namely, depending on
the relative values of $k$ and $z$, we have:

\noindent $\bullet$ Wigner regime $(k\ll |z|)$
\begin{equation}
   \sigma^{\rm in.} \simeq \frac{4\pi}{k} \, e^{-2k\tilde{\beta}}
   \left( \frac{z'}{|z|^2} + \tilde{\beta}\right)
    \ \xrightarrow[\ k\rightarrow 0\ ]\ \ \frac{4\pi}{k} \beta .
    \label{eq:wigner}
\end{equation}
\noindent $\bullet$ NTR regime $(|z|\ll k)$
\begin{equation}
    \sigma^{\rm in.} \simeq 4\pi \: e^{-2k\tilde{\beta}} \frac{z'}{k^3}
    \ \ \xrightarrow[\ k\rightarrow 0\ ]\\ \ 4\pi\frac{z'}{k^3} \;.
    \label{eq:RTS} 
\end{equation}
In Eq.~(\ref{eq:wigner}), the imaginary component of the
background scattering length, $\tilde{\beta}$, is augmented by a
contribution from the resonance at $k=-z$. This is also true for the
real component; writing $S_{ii}=e^{2i\delta}$ with $\delta \approx
-ka$ as $k\rightarrow 0$, and expanding (\ref{eq:S_ee}) to the leading
order in $k$, so that $1-2ika  \approx 1
-2ik(\tilde{a} -\frac{i}{z})$, we identify $a=\tilde{a}
-\frac{i}{z}$. Separating the real and imaginary components, we have
\begin{eqnarray}
  \alpha & = & \tilde{\alpha} + \frac{z''}{|z|^2} \;, \label{eq:alpha_e} \\
  \beta  & = & \tilde{\beta}  + \frac{z'}{|z|^2} \label{eq:beta_e} \;.
\end{eqnarray}
If $z$ is very close to zero, the effect on the Wigner's regime can be
significant, with large increases of the cross sections from their
background values.

\begin{figure}[t]
\includegraphics[clip,width=.85\linewidth]{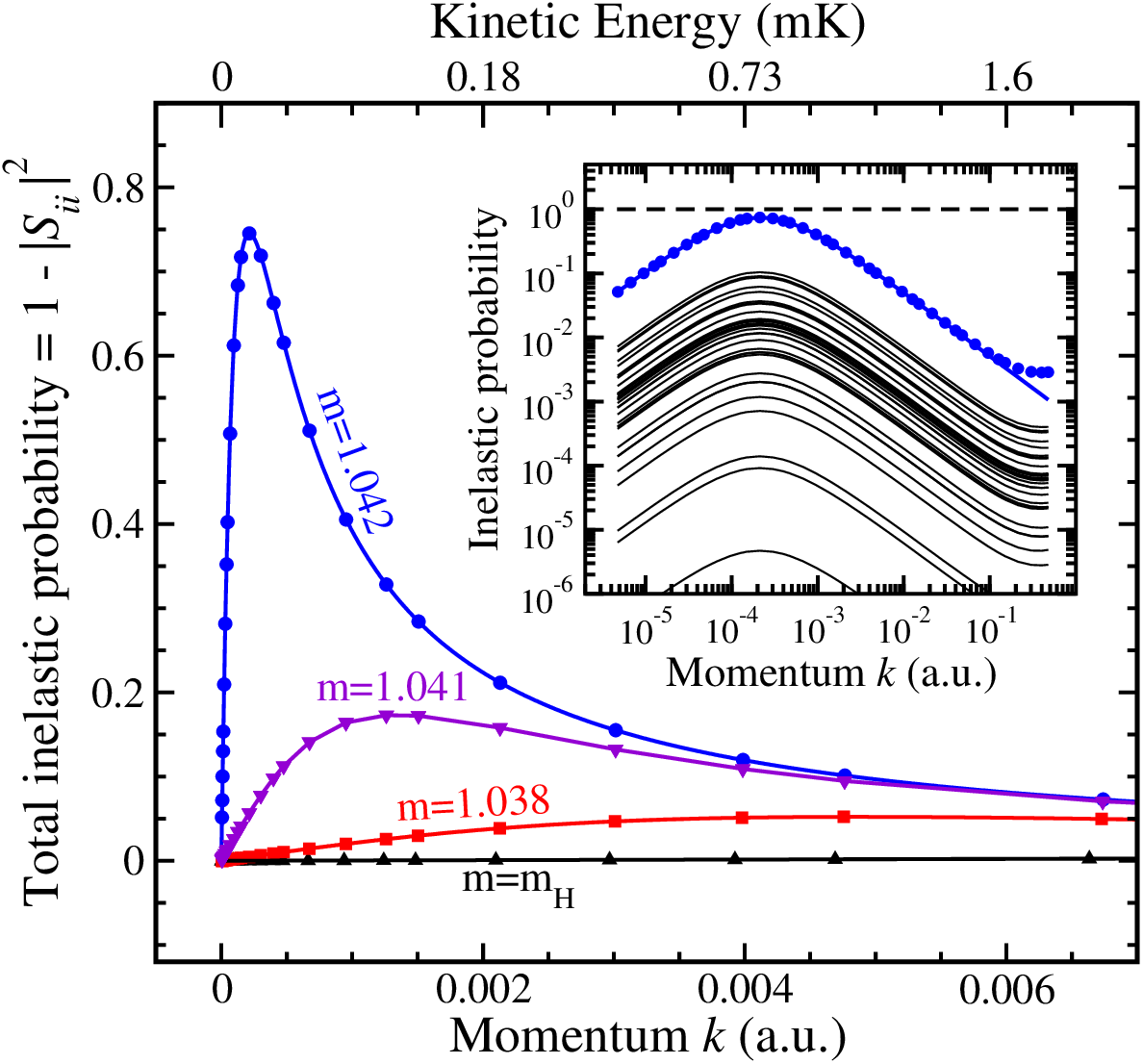}
\caption{\label{fig:smat} Momentum (lower scale) and energy (top
  scale) dependence of the total inelastic probability $\sigma^{\rm
    in.} k^2 / \pi = 1 -|S_{ii}|^2$ for different values of $m$ on
  linear scales: symbols for the full (exact) computation, and lines
  for the analytical formula (\ref{eq:sigma_in-full}). Inset: results
  for $m=1.042$\,u on logarithmic scales, showing the many individual
  state-to-state inelastic probabilities (thin black lines).  The
  dashed horizontal line marks the unitarity limit.}
\end{figure}

Fig.~\ref{fig:smat} shows the inelastic probability $1-|S_{ii}|^2$ for
various masses ranging from the true mass of hydrogen ($m=m_{\rm H}$)
to $m=1.042$\,u.  The linear scale accentuates the asymmetrical
profile of the inelastic probability near threshold, and the good
agreement between the numerical results and the analytical expression
(\ref{eq:sigma_in-full}). The values of $\tilde{\beta}$ and $z$ can be
obtained by fitting the calculated $\sigma^{\rm in.}$ to
Eq.~(\ref{eq:sigma_in-full}).  The inset shows the results for
$m=1.042$\,u on a log-log scale, together with the maximum possible
value of unity (corresponding to the unitarity limit, when
$S_{ii}=0$).  It reveals the simplicity of the low energy behavior in
the presence of NTR; the Wigner power-law scaling for $k\ll |z|$, a
different power-law for $k\gg|z|$, and the transition between the two
regimes taking place near $k\approx|z|$.  The inset also illustrates
the deviation of the analytical expression from numerical results at
larger $k$, where $\delta\simeq -ka$ ceases to be valid.  Finally, the
state-to-state results for all final channels are also shown,
exhibiting the same resonant behavior. It also emphasizes that although
Eq.~(\ref{eq:sigma_in-full}) is rather simple, the coupled channel
computations are quite complex, involving several hundreds of channels
(open and closed).

For the sake of completeness, we apply the same approach 
to the elastic cross section (\ref{eq:sigma-el-S-matrix}) and obtain
\begin{eqnarray}
  \sigma^{\rm el.}
  & = & \frac{\pi}{k^2} \frac{2 e^{-2k\tilde{\beta}}}{|z+k|^2}
        \left\{ (k^2+|z|^2) \cosh 2k\tilde{\beta} 
              + 2 k z'' \sin 2k\tilde{\alpha} \right. \nonumber \\
  &   &  \makebox[.25in]{ } 
  \left. {} + 2 k z' \sinh 2k\tilde{\beta}
        + (k^2-|z|^2) \cos 2k\tilde{\alpha}  \right\}   \;.
\label{eq:sigma-el-full}
\end{eqnarray} 
Again, as $k\rightarrow 0$,  two different behaviors are found:

\noindent $\bullet$ Wigner regime $(k\ll |z|)$
\begin{eqnarray}
    \sigma^{\rm el.} & \simeq & 4\pi \, e^{-2k\tilde{\beta}} 
   \left\{\frac{1}{|z|^2} + (\tilde{\alpha}^2 + \tilde{\beta}^2) 
   +\frac{2}{|z|^2} (z''\tilde{\alpha} + z'\tilde{\beta}) \right\} \nonumber \\ & 
    = & 4\pi \: e^{-2k\tilde{\beta}}  |a|^2 
        \ \xrightarrow[\ k\rightarrow 0\ ]\ \  4\pi |a|^2 \;.
    \label{eq:el-wigner}
\end{eqnarray}
\noindent $\bullet$  NTR regime $(|z|\ll k)$
\begin{eqnarray}
   \sigma^{\rm el.}  &  \simeq  &  \frac{4\pi}{k^2} \, e^{-2k\tilde{\beta}} 
                         \left\{1 + |z|^2 (\tilde{\alpha}^2 + \tilde{\beta}^2) 
                      +2 (z''\tilde{\alpha} + z'\tilde{\beta}) \right\} \nonumber \\
     & \approx & \frac{4\pi}{k^2} \, e^{-2k\tilde{\beta}}
     \ \ \xrightarrow[\ k\rightarrow 0\ ]\ \ \ \frac{4\pi}{k^2} \;.
    \label{eq:el-RTS} 
\end{eqnarray}
Above, we expressed explicitly $a=\alpha-i\beta$ in term of
Eqs.~(\ref{eq:alpha_e}) and (\ref{eq:beta_e}). These results for the
elastic cross section are well known and simply understood in term of
the phase shift variation. For $k\rightarrow 0$, $\delta\approx -ka$
leads to $S=e^{2i\delta}\approx 1-2ika$ and
Eq.~(\ref{eq:sigma-el-S-matrix}) gives $\sigma^{\rm el.} \approx 4\pi
|a|^2$. However, when the potential gains a new $s$-wave bound state
near a resonance, the phase shift abruptly reaches $\pm\pi/2$ (or $ka$
quickly reaches large values even though $k$ is small), giving then
$S\approx e^{\pm i\pi}=-1$ and $\sigma^{\rm el.} \approx 4\pi /k^2$.

We conduct our analysis in the complex $k$-plane rather than the energy's,
since the complex energies associated with both the zero ($k=z$) and
the pole ($k=-z$) are the identical;
\begin{equation} 
   E_z = \frac{\hbar^2 (\pm z)^2}{2\mu}
     \equiv \varepsilon_z - i\hbar\frac{\Gamma}{2} \;,
\end{equation}
with $\varepsilon_z = \frac{\hbar^2}{2\mu}(z'^2-z''^2)$ and $\Gamma =
\frac{2\hbar}{\mu} z'z''$ remaining unchanged for $\pm z$. As shown in
Eq.~(\ref{eq:RTS}), we must have $z'>0$ for a physical cross section,
and thus the decay rate $\Gamma$ of the vdW-complex is positive
(physical) in the quasi-bound case ($z''>0$), and negative (hence,
unphysical) in the anti-bound case ($z''<0$). For an $s$-wave NTR,
inelastic scattering alone cannot distinguish between shallow bound
and anti-bound (virtual) states, since $\sigma^{\mathrm{in.}}(k)$
depends on $(z'')^2$ (see Eq.~(\ref{eq:sigma_in-full})).

In conclusion, we found that a near threshold resonance (NTR) 
splits the $s$-wave low energy domain in two distinct regimes, with the
total inelastic cross section gradually transitioning from the well
known $k^{-1}$ Wigner regime into the $k^{-3}$ NTR regime. We derived
simple analytical expressions for both elastic and total inelastic
cross sections as a function of the position of the zero/pole of the
$S$-matrix, and found very good agreement with values obtained
numerically.  This $s$-wave $k^{-3}$ NTR behavior is a general
feature, but its presence can be masked by higher partial wave
contributions.  Although previous work hinted at such effect using 
mass-scaling \cite{Bodo:JPB:2004,Alex:PRA2010:He-H2} or external fields
\cite{JMH:PRL:2009:Feshbach}, the new universal NTR $k^{-3}$ behavior
shown in Eq.~(\ref{eq:sigma_in-full}) remained elusive until now. We
revealed the effect of NTR in a benchmark atom-diatom reactive
scattering by mass-scaling, but it is should appear in any system with
a zero-energy resonance, such as in photoassociation
\cite{Crubellier:JPB:2006,FOPA}, 
collision \cite{Cs2:resonance}, or spin-relaxation in ultracold atomic 
samples \cite{Cs2:spin}. The
modified $k^{-3}$ inelastic cross section will impact not only the 
interpretation of experiments such as \cite{narevicius-2012}, but also
theories developed to account for resonances in ultracold molecular systems
\cite{Bohn:PRA87:2013}.

This work was partially supported by the US Department of Energy,
Office of Basic Energy Sciences (SG), the Air Force Office of
Scientific Research MURI (IS) and the National Science Foundation
Grant No. PHY 1101254 (RC). 


\begin{thebibliography}{99}
\bibitem{RMP-FR}
  C. Chin, R. Grimm, P. Julienne, and E. Tiesinga,
  Rev. Mod. Phys. {\bf 82}, 1225 (2010).
\bibitem{paper-JILA}
  M. H. G. de Miranda, A. Chotia, B. Neyenhuis, D. Wang, G.~Qu\'em\'ener,
  S. Ospelkaus, J. L. Bohn, J. Ye, and D. S. Jin,
  Nature Phys. {\bf 7}, 502 (2011).
\bibitem{Jason-PRL}
  J. N. Byrd, J. A. Montgomery, Jr., and R. C\^ot\'e, 
  Phys. Rev. Lett. {\bf 109}, 083003 (2012).
\bibitem{RMP-bose}
  F. Dalfovo, S. Giorgini, L. P. Pitaevskii, and S. Stringari,
  Rev. Mod. Phys. \textbf{71}, 463 (1999);
  A. J. Leggett,
  Rev. Mod. Phys. \textbf{73}, 307 (2001).
\bibitem{RMP-fermi}
  S. Giorgini, L. P. Pitaevskii, and S. Stringari,
  Rev. Mod. Phys. \textbf{80}, 1215 (2008).
\bibitem{efimov}
  T. Kraemer, M. Mark, P. Waldburger, J. G. Danzl, C. Chin, B. Engeser,
  A. D. Lange,  K. Pilch, A. Jaakkola, H.-C. N\"agerl, and R. Grimm,
  Nature {\bf 440}, 315 (2006).
\bibitem{carr2009} 
  L. Carr, D. DeMille, R. Krems, and J. Ye,
  New J. Phys. {\bf 11}, 055049 (2009).
\bibitem{dulieu2011}
  O. Dulieu, R. Krems, M. Weidem\"uller, and S. Willitsch,
  Phys. Chem. Chem. Phys. {\bf 13}, 18703 (2011).

\bibitem{mol-papers}
  D. S. Jin and J. Ye,
  Chem. Rev. {\bf 112}, 4801 (2012), and references therein.
\bibitem{sawyer2011}
  B. C. Sawyer, B. K. Stuhl, M. Yeo, T. V. Tscherbul, M. T. Hummon, Y. Xia, 
  J. Klos, D. Patterson, J. M. Doyle, and J. Ye,
  Phys. Chem. Chem. Phys. {\bf 13}, 19059 (2011).
\bibitem{quemener2012}
  G. Qu\'em\'ener and P. Julienne,
  Chem. Rev. {\bf 112}, 4949 (2012).
\bibitem{h2co-prl-2012}
  S. Chefdeville, T. Stoecklin, A. Bergeat,
  K. M. Hickson, C. Naulin, and  M. Costes,
  Phys. Rev. Lett. \textbf{109}, 023201 (2012).
\bibitem{shd-prl-2012}
  M. Lara, S. Chefdeville, K. M. Hickson, A. Bergeat, C. Naulin, J.-M. Launay, 
  and M. Costes,
  Phys. Rev. Lett. \textbf{109}, 133201 (2012).
\bibitem{narevicius-2012}
  A. B. Henson, S. Gersten, Y. Shagam, J. Narevicius, and E.~Narevicius,
  Science \textbf{338}, 234 (2012). 
\bibitem{Bohn:PRA87:2013}
  M. Mayle, G. Qu\'em\'ener, B. P. Ruzic, and J. L. Bohn,
  Phys. Rev. A {\bf 87}, 012709 (2013).
\bibitem{wigner}
  E. P. Wigner,
  Phys. Rev. \textbf{73}, 1002 (1948).

\bibitem{hossein}
H. R. Sadeghpour, J. L. Bohn, M. J. Cavagnero,
B. D. Esry, I. I. Fabrikant, J. H. Macek, A. R. P. Rau,
J. Phys. B \textbf{33}, R93 (2000).

\bibitem{Bethe:PR:1935}
  H. A. Bethe, 
  Phys. Rev. \textbf{47}, 747 (1935).
\bibitem{fano}
  U. Fano,
  Phys. Rev. \textbf{124}, 1866 (1961).
\bibitem{ClH2:Science:2008}
  E. Garand, J. Zhou, D. E. Manolopoulos, M. H. Alexander, and D. M. Neumark, 
  Science \textbf{319}, 72 (2008).
\bibitem{Bala:H2Cl:2012}
  N. Balakrishnan, J. Chem. Sci. {\bf 124}, 311 (2012).   
\bibitem{Bodo:JPB:2004}
  E. Bodo, F. A. Gianturco,  N. Balakrishnan, and A. Dalgarno,
  J.~Phys. B \textbf{37}, 3641 (2004).
\bibitem{JMH:JCP:2007}
  M. T. Cvitas,  P. Soldan,  J. M. Hutson,  P. Honvault, and \mbox{J.-M. Launay},
  J. Chem. Phys. \textbf{127}, 074302 (2007).
\bibitem{alex-bala}
    N. Balakrishnan, R. C. Forrey, and A. Dalgarno,
  Phys. Rev. Lett. \textbf{80}, 3224 (1998).
\bibitem{bala-cplett}
  N. Balakrishnan, V. Kharchenko, R. C. Forrey, and A. Dalgarno,
  Chem. Phys. Lett. \textbf{280}, 5 (1997).
\bibitem{ABC:CPC:2000}
  D. Skouteris, J. F. Castillo, D. E. Manolopoulos,
  Comp. Phys. Comm. \textbf{133}, 128 (2000).
\bibitem{BWpes:JCP:2000}
  W. Bian and H.-J. Werner,
  J. Chem. Phys. \textbf{112}, 220 (2000).



\bibitem{Alex:PRA2010:He-H2}
  J. L. Nolte, B. H. Yang, P. C. Stancil, T.-G. Lee, N. Balakrishnan,
  R. C. Forrey,  and A. Dalgarno,
  Phys. Rev. A {\bf 81}, 014701 (2010).
 
\bibitem{JMH:PRL:2009:Feshbach}
  J. M. Hutson, M. Beyene, and M. L. Gonz\'alez-Mart\'\i{}nez, 
  Phys. Rev. Lett. \textbf{103}, 163201 (2009). 
  

\bibitem{Crubellier:JPB:2006}
  A. Crubellier and E. Luc-Koenig,
  J. Phys. B \textbf{39}, 1417 (2006).

   
\bibitem{FOPA}
  P. Pellegrini, M. Gacesa, and R. C\^ot\'e, 
  Phys. Rev. Lett. {\bf 101}, 053201 (2008).
\bibitem{Cs2:resonance}
  M. Arndt, M. Ben Dahan, D. Gu\'ery-Odelin, M. W. Reynolds, and J. Dalibard, 
  Phys. Rev. Lett. {\bf 79}, 625 (1997). 
   
\bibitem{Cs2:spin}
  J. S\"oding, D. Gu\'ery-Odelin, P. Desbiolles, G. Ferrari, and J.~Dalibard, 
  Phys. Rev. Lett. {\bf 80}, 1869 (1998).

\end{thebibliography}
\end{document}